\documentclass[aps,pra,reprint, showpacs, superscriptaddress]{revtex4-1}

\usepackage{epsfig,graphics,graphicx}
\usepackage{dcolumn}
\usepackage{bm}
\usepackage{amsmath,amssymb,amsthm}
\usepackage{centernot}
\usepackage{bbold}
\usepackage{soul}
\usepackage{mathtools}
\usepackage{fullpage}
\usepackage{units}
\usepackage[usenames,dvipsnames]{xcolor}
\usepackage{braket}
\usepackage{indentfirst}
\usepackage{lipsum}
\usepackage[export]{adjustbox}

\usepackage[utf8]{inputenc}
\usepackage[T1]{fontenc}

\usepackage{dsfont}

\usepackage{mathptmx}
\usepackage{amssymb}
\usepackage{amsmath}
\usepackage{latexsym}
\usepackage{amsfonts}

\usepackage{hyperref}
\hypersetup{pdfpagemode=UseNone}

\newtheorem{theorem}{Theorem}
\newtheorem{definition}[theorem]{Definition}

\newtheorem{lemma}[theorem]{Lemma}
\newtheorem{proposition}[theorem]{Proposition}

\newcounter{rem}
\newcommand{\mc}[1]{\mathcal{#1}}

\DeclarePairedDelimiter\floor{\lfloor}{\rfloor}

\def\>{\rangle}
\def\<{\langle}

\renewcommand{\rho}{\varrho}
\DeclareMathOperator{\pr}{Pr}

\def\ii{{\rm i}}

\def\textbf#1{{\bf #1}}

\def\beq{\begin{equation}}
\def\eeq{\end{equation}}
\def\beqa{\begin{eqnarray}}
\def\eeqa{\end{eqnarray}}
\def\eea{\end{array}}
\def\bea{\begin{array}}
\newcommand{\bei}{\begin{itemize}}
\newcommand{\eei}{\end{itemize}}
\newcommand{\bee}{\begin{enumerate}}
\newcommand{\eee}{\end{enumerate}}
\def\bep{\begin{proposition}}
\def\eep{\end{proposition}}
\def\bel{\begin{lemma}}
\def\eel{\end{lemma}}
\def\bet{\begin{theorem}}
\def\eet{\end{theorem}}
\def\bed{\begin{definition}}
\def\eed{\end{definition}}
\usepackage{soul}
	
\begin{document}

\title{Entropic Uncertainty Relations and the Quantum-to-Classical transition}

\author{Isadora Veeren}
\email{veeren@cbpf.br}
\affiliation{Centro Brasileiro de Pesquisas Físicas,\\ Rua Dr. Xavier Sigaud, 150, Rio de Janeiro, RJ, Brasil}

\author{Fernando de Melo}
\email{fmelo@cbpf.br}
\affiliation{Centro Brasileiro de Pesquisas Físicas,\\ Rua Dr. Xavier Sigaud, 150, Rio de Janeiro, RJ, Brasil}

\date{\today}


\begin{abstract}
Our knowledge of quantum mechanics can satisfactorily describe simple, microscopic systems, but is yet to explain the macroscopic everyday phenomena we observe. Here we aim to shed some light on the quantum-to-classical transition as seen through the analysis of uncertainty relations. We employ entropic uncertainty relations to show that it is only by the inclusion of imprecision in our model of macroscopic measurements that we can prepare a system with two simultaneously well-defined quantities, even if their associated observables do not commute.   We also establish how the precision of measurements must increase in order to keep quantum properties, a desirable feature for large quantum computers.
\end{abstract}


\maketitle

\section{Introduction}

By the current scientific point of view the world is quantum. Yet, a range of  quantum phenomena, such as  quantum tunneling~\cite{MandelstamTunneling} and entanglement among quantum particles~\cite{EPR, schroedinger1935, HorodeckiRev},   are not observed in our daily life.

The issue of translating quantum mechanics to our everyday macroscopic world has been discussed since  early stages of the field. When confronted with the subject, Schrödinger presented his ``cat paradox''~\cite{SchrodingerCat}, illustrating the weird scenarios we end up with when we simply force quantum mechanics into macroscopic descriptions. In the last century, the decoherence program lead to a partial understanding of the quantum-to-classical transition~\cite{caldeira, zurek2003,schlosshauer}, taking into account that quantum systems cannot be completely isolated. Recent experiments, however, have been pushing forward the size of systems that can exhibit genuine quantum features~\cite{Eibenberger, connell, riedinger}, and as such they  bring back the Schrödinger's cat discussion to the forefront of the physics agenda.

Moreover, in the flourishing field of quantum computation, the quantum-to-classical transition stopped from being an exclusively foundational question to become also an applied one. With the number of qubits quickly increasing in quantum computers~\cite{ibm, ionq, google}, we must address how to preserve the quantum features  which eventually will allow  macroscopic quantum computers  to tackle real world problems in an efficient way.
 
In recent years, with the development of the quantum information field, a coarse graining argument is being advanced in order to explain the quantum-to-classical transition even for closed systems~\cite{mermin1980,poulin2005,caslavLG,Raeisi2011,Wang2013,Jeong2014,Park2014,cris2017,pedrinho,oleg,cris2019}. The coarse graining approach can be seen as an extension of the decoherence theory~\cite{ZurekRMP} where we employ generalized  subsystems~\cite{alicki2009,oleg2020a}. 

The main idea of the coarse graining method is that the classical behaviour might emerge depending on the resolution one describes the system. For highly precise measurements one can observe genuine quantum features. While when we only have coarsed access to the system, its quantum signatures might  vanish and an effective classical description emerges.
In references~\cite{mermin1980,Jeong2014} it was shown that imprecise measurements might turn violations of Bell inequalities impossible to be observed. In the same direction, the vanishing of superpositions~\cite{Wang2013,Park2014},  quantum entanglement~\cite{Raeisi2011,pedrinho}, and violation of Leggett-Garg inequalities~\cite{caslavLG}, were all shown to happen due to a coarse-grained description of the quantum system.

With these motivations in mind, the goal of this work is to further investigate the preparation of quantum macroscopic systems, a striking distinguishing feature between quantum and classical structures. Both descriptions adopt observables to characterize properties of a system, but quantum properties must, additionally, abide by uncertainty relations. Here we employ preparation uncertainty relations, in spite of error-disturbance inequalities~\cite{Ozawa03,Busch13}, to analyse what are the necessary conditions in order to prepare a quantum system with two well-defined properties, even when to these properties are associated non-commuting observables.


\section{Preparation Uncertainty Relations} 
One of the foundational results of quantum theory is the Heisenberg Uncertainty Relation (HUR)~\cite{HeisenbergUncertainty}. Introduced already in 1927, in its more common form~\cite{Robertson} it reads:
\begin{equation} 
\label{robertson}
\Delta(A|\Psi)\; \Delta(B|\Psi) \geq \frac{1}{2} | \< \left[ A,B \right]\>_\Psi |.
\end{equation}
That is,  given an assigned Hilbert space $\mc{H}$ with a preparation $\ket{\Psi}\in\mc{H}$, and two physical properties with associated observables $A$ and $B$ acting on $\mc{H}$,  the product of the variances $\Delta(A|\Psi)$ and $\Delta(B|\Psi)$ associated with the properties' measurement statistics --- where  $\Delta(P|\Psi):=\sqrt{\<P^2\>_\Psi- \<P\>^2_\Psi}$ with $P\in\{A,B\}$ and $\<P\>_\Psi:=\<\Psi|P|\Psi\>$ ---, is lower bounded by half of the absolute value of the expectation of their commutator, $[A,B]=AB-BA$.  Physically, the HUR poses a restriction on the preparation of a system:  properties $A$ and $B$ can only be simultaneously well-defined for a preparation $\ket{\Psi}$, if $\ket{\Psi}$ is a common eigenstate of $A$ and $B$.

Given the HUR formulation, Eq.~\eqref{robertson}, when trying to understand the emergence of  classical behaviour, the focus was on the commutation relation. Already in 1929, John von Neumann suggested that the actual classical observables related to position and momentum are commuting versions of the ``true'' quantum observables~\cite{neumann29}. When dealing with the thermodynamic limit of finite-dimensional observables, the lore goes as follows: consider, for instance, the (dimensionless) observables associated to the magnetization in three orthogonal directions, namely:
\begin{align}
X_N = \frac{1}{N}\sum_{i=1}^N \frac{\sigma_{x}^{(i)}}{2},&& Y_N = \frac{1}{N}\sum_{i=1}^N \frac{\sigma_{y}^{(i)}}{2},&& Z_N = \frac{1}{N}\sum_{i=1}^N \frac{\sigma_{z}^{(i)}}{2}.
\label{magnetization}
\end{align}
Here $N$ is the total number of spin-1/2 particles, and $\sigma_k^{(i)}$ is the $k$-th Pauli matrix, with $k\in\{x,y,z\}$, acting on the $i$-th spin. Taking two of these observables, say $X_N$ and $Z_N$, we have $[X_N, Z_N]=-\ii Y_N/N$. As $||Y_N||=1$, when $N$ goes to infinity $\lim_{N\rightarrow \infty}||[X_N, Z_N]||=0$. One may be tempted to say that it is then possible to prepare a state $\ket{\Psi}$ with simultaneously well defined magnetization in $x$ and $z$ directions for large systems. However,  that is not the case, as $X_N$ and $Z_N$ do not share any common eigenvector for any (finite) value of $N$.

The above misconceptions are due to shortcomings of the Heisenberg uncertainty relation~\cite{Deutsch1983}. Most prominently, the HUR is sensitive to rescaling of the observables. By changing the eigenvalues associated with the observables, we can make the lower bound in Eq.~\eqref{robertson} to assume any positive value. All that this uncertainty relation indicates is that the lower bound is either zero or non-zero. Moreover, for a pair of observables that are not infinite-dimensional canonically conjugated variables, the right-hand-side of Eq.~\eqref{robertson} is state dependent and as such may be not so useful. In the magnetization case, take for instance $\ket{\Psi}$ as  an eigenvector of $Z_N$. Both the right-hand-side and the left-hand-side of Eq.~\eqref{robertson} go to zero, and nothing can be said about $\Delta(X_N|\Psi)$.

With the advent of quantum information science, Entropic Uncertainty Relations (EUR) were introduced to address  HUR' shortcomings~\cite{Deutsch1983,Kraus,Maassen,review1,review2,review3}. Such relations use entropies as measures of uncertainty, and imply the Robertson uncertainty principle~\cite{Birula1975}.
For two given observables $A$ and $B$, with eigenvectors $\{\ket{a_j}\}$ and $\{\ket{b_k}\}$, the EUR based on Shannon's entropy reads:
\begin{equation} \label{EntropicUncertainty}
	H(A|\Psi) + H(B|\Psi) \geq -2 \log \max_{j,k} |\braket{a_{j}|b_{k}}|,
\end{equation}
where $H(A|\Psi) = - \sum_{j}|\braket{\Psi|a_{j}}|^{2} \log |\braket{\Psi|a_{j}}|^{2}$ is the entropy associated with the measurement of $A$ on the state $\ket{\Psi}$, and similarly for $H(B|\Psi)$. 

Much like Heisenberg's uncertainty principle, the EUR~\eqref{EntropicUncertainty} sets a lower bound for how  well-defined the properties $A$ and $B$ can simultaneously be in a preparation $\ket{\Psi}$. Notice, however, that in this case the lower bound  is state independent, and it also does not depend on the observables eigenvalues. These features make the entropic uncertainty relations the suitable relation to analyze the quantum-to-classical transition for physical properties and preparations. In a classical regime where we can prepare a system with two well-defined properties, one would expect the sum of entropies to vanish as the system increases.

Nevertheless, back to the magnetization observables, it is simple to show that 
\begin{equation}
\label{boundN}
H(X_N|\Psi) + H(Z_N|\Psi)   \geq N.
\end{equation}   
The lower bound now, contrary to what is suggested by the HUR case, increases with $N$. A classical behavior is thus not directly obtained by simply increasing  the system size.

For clarity, in the rest of the article we will concentrate on the preparation of a macroscopic system with well-defined magnetization in two orthogonal directions.


\section{Macroscopic Preparations}  As expected from the bosonic case~\cite{glauber}, spin-coherent states~\cite{arecchi} either in the $x$ or in the $z$ direction saturate inequality~\eqref{boundN}. More concretely, if we define the Pauli eigenvectors as $\sigma_{z}\ket{s} =(-1)^s\ket{s}$, with $s\in\{0,1\}$, then the states in the set
$\{\ket{0}^{\otimes N},\ket{1}^{\otimes N},\ket{+}^{\otimes N},\ket{-}^{\otimes N}\}$,
where $\ket{\pm} = (\ket{0}\pm\ket{1})/\sqrt{2}$, are spin-coherent states that saturate the bound~\eqref{boundN}. 

For generic spin-coherent states one can evaluate the sum of entropies in~\eqref{boundN}. Let $\ket{\Psi_1}=\sqrt{p}\ket{0}+ e^{\ii \phi}\sqrt{1-p}\ket{1}$, with $p\in[0,1]$ and $\phi\in [0,2\pi[$, be the state of a single spin,  and
\beq
\ket{\Psi_N }=\ket{\Psi_1}^{\otimes N}
\label{eq:coh}
\eeq
be the state of the full $N$ spin-coherent state. The entropy associated with the measurement in the $z$ direction is given by:
\begin{align}
H(Z_N|\Psi_N)&= - \sum_{k=0} {N \choose k} p^{N-k}(1-p)^{k}\log p^{N-k}(1-p)^{k},\nonumber\\
&=N\, h(p),
\end{align}
where $h(p)= -p \log p - (1-p)\log (1-p)$ is  Shannon's binary entropy. Writing $\ket{\Psi_1}$ in the basis of eigenvectors of $\sigma_{x}$, $\ket{\Psi_1}= [(\sqrt{p}+e^{\ii \phi}\sqrt{1-p})\ket{+}+(\sqrt{p}-e^{\ii \phi}\sqrt{1-p})\ket{-}]/\sqrt{2}$, a similar calculation leads to $H(X_N|\Psi_N)=N\, h(q)$, where $q= \frac{1}{2} +\sqrt{p(1-p)}\cos \phi$ is the probability of projecting $\ket{\Psi_1}$ onto $\ket{+}$. Putting these together, for generic spin-coherent state we have
\beq
H(X_N|\Psi_N)+H(Z_N|\Psi_N)= N[h(p)+h(q)],
\label{eq:boundCoh}
\eeq
which grows linearly with $N$ and, in the $x-z$ plane, saturates~\eqref{boundN} for $p\in\{0,1/2,1\}$, as mentioned before.

Besides being the analog of coherent states  for spins~\cite{arecchi}, the states of the form in~\eqref{eq:coh} play an important role in the quantum-to-classical transition. In the theory of quantum darwinism~\cite{zurek,brandao,sheilla} such states are responsible for the redundant encoding of a system's property, allowing for different observers to agree on the value of such a property. However, like demonstrated by Eq.~\eqref{eq:boundCoh}, in Ref.~\cite{caslavLG} the mere use of coherent states is shown to not be sufficient for a classical behavior -- signaled there by the no violation of Leggett-Garg inequality~\cite{LG} -- to emerge. A coarse-grained measurement is also required.

\section{Macroscopic measurements: degeneracy} In order to obtain the results in~\eqref{boundN} and~\eqref{eq:boundCoh} we assumed that each eigenvector of $X_N$ and $Z_N$ could be independently measured. This presumes the capacity of individual spin measurement. Such a level of control is nor expected neither desirable in macroscopic systems --  the measurement of, say, $Z_N$ would entail a POVM with $2^N$ outcomes.

Macroscopic quantities, however, are usually insensitive to small differences in the microscopic systems, i.e., their associated observables are highly degenerate. When preparing a macroscopic system with a given magnetization, we are often more interested in the total spin than on each individual spin value. Making the degeneracy of the magnetization observables in the $x$ and $z$ directions explicit, we write the total magnetization observables as:
\begin{align}
\label{eq:totalspin}
\tilde{X}_N = \frac{1}{N}\sum_{j_x=-N/2}^{N/2} j_x \Pi_x(j_x),\;&&\tilde{Z}_N = \frac{1}{N}\sum_{j_z=-N/2}^{N/2} j_z \Pi_z(j_z),
\end{align}
where $\Pi_k(j_k)$, with $k\in\{x,z\}$ is the projector onto the subspace of $j_k$ total spin in direction $k$. The exponential number of outcomes mentioned above turns now into $N+1$ possibilities for each direction.

Profiting from the already established form of spin coherent states, Eq.~\eqref{eq:coh}, it is simple to realize that the probability of obtaining the outcome $j_z/N$ is given by

\beq
\pr(j_z|\Psi_N)= {N \choose \frac{N}{2} +j_z}p^{\frac{N}{2} +j_z}(1-p)^{\frac{N}{2} -j_z}.
\eeq
This leads to a binomial distribution for the eigenvalues $j_z/N$ of $\tilde{Z}_N$. Such a distribution has mean $\<\tilde{Z}_N\>_{\Psi_N} = p-1/2$,  and standard deviation $ \Delta^{2}(\tilde{Z}_N|\Psi_N)=p(1-p)/N$. The distribution concentrates around the mean as $1/\sqrt{N}$. However, as the number of outcomes grows linearly with $N$, the entropy of such a distribution does not vanish for large systems. In fact,  in the limit $N \gg 1$ the entropy $H(\tilde{Z}_N|\Psi_{N})=- \sum_{j_z=-N/2}^{N/2} \pr(j_z|\Psi_N) \log \pr(j_z|\Psi_N)$ is approximately given by $\frac{1}{2} \log 2 \pi e Np(1-p)$ (where we used the continuous limit for the probability distribution~\cite{feller} and for the entropy function). 

A totally analogous derivation can be followed for $\tilde{X}_N$, and in the macroscopic limit we get:
\begin{align}
H(\tilde{X}_N|\Psi_{N})+&H(\tilde{Z}_N|\Psi_{N})\approxeq \nonumber \\
& \log N + \frac{1}{2} \log 4 \pi^2 e^2  pq(1-p)(1-q).
\end{align}
Although slower than in Eq.\eqref{eq:boundCoh}, even when taking into account the degeneracy of macroscopic quantities,  the sum of entropies still grows with the system size $N$.

\section{Macroscopic measurements: division into bins}  The above description of macroscopic observables is still not realistic. 
 As the number of outcomes is $N+1$,  measuring the total magnetization in one direction of a system composed of $10^{23}$ spins requires an inconceivable precision. 
 
One last ingredient has then to be observed. Typical measurement apparatuses have fixed precision for different system sizes. The measurement of magnetization in usual Nuclear Magnetic Resonance (NMR), for instance, uses the same apparatus for sample sizes around $10^{13}$ molecules. Moreover, the experiment, which actually measures frequencies, has  precision of $0.5 Hz$ for frequencies around $500 MHz$ (the Hydrogen Larmor frequency in a magnetic field of $11T$)~\cite{NMRbook}.  All that means that our model must have a number of outcomes that is independent of the system size, i.e., a fixed number of bins, and that all the magnetization values within a bin are integrated to correspond the bin value.

To assimilate the notion of imprecision in our description,  like in~\cite{caslavLG,poulin2005}, we will group neighboring results under a same bin of width $\delta$, which we suppose to be the same for both $x$ and $z$ directions. In this way, we incorporate our inability to distinguish between nearby outcomes of $\tilde{X}_N$ and $\tilde{Z}_N$. Instead of evaluating the probability of a state having a total magnetization $j_z/N$, we will evaluate their probability of belonging to the interval $[j_z/N-\delta/2,j_z/N+\delta/2[$. 

Notice that the number of bins, $N_{b}$, is related to the bin width $\delta$ by $N_{b} =N/\delta$. Thus, in terms of the number of bins $N_{b}$, the $n$-th bin will cover magnetizations in the interval $	[ -\frac{1}{2} + \frac{n-1}{N_{b}}, -\frac{1}{2} + \frac{n}{N_{b}} [$, with $n\in\{1,\ldots, N_b\}$ --- magnetization $1/2$ is included in the last bin. 

To make  this more realist setup explicit, the magnetization observables are now written as follows:
\begin{align}
\label{eq:bintotalspin}
X^\prime_N = \frac{1}{N}\sum_{n_x=1}^{N_b} j_{n_x}\Bigg( \sum_{\frac{j_x}{N}\in [ -\frac{1}{2} + \frac{n_x-1}{N_{b}}, -\frac{1}{2} + \frac{n_x}{N_{b}} [}  \Pi_x(j_x)\Bigg),\nonumber\\ 
Z^\prime_N = \frac{1}{N}\sum_{n_z=1}^{N_b} j_{n_z}\Bigg(\sum_{\frac{j_z}{N}\in [ -\frac{1}{2} + \frac{n_z-1}{N_{b}}, -\frac{1}{2} + \frac{n_z}{N_{b}} [} \Pi_z(j_z)\Bigg).
\end{align}
Above, $j_{n_k}/N$, with $k\in\{x,z\}$, is  the  magnetization eigenvalue associated with the bin $n_k$ of  direction $k$. As the entropic uncertainty relations do not depend on the eigenvalues, we don't need to specify them explicitly. 
More importantly, notice that for both directions the number of outcomes is $N_b$, which it is fixed by the measurement apparatus precision, and thus independent of the system size $N$.

For spin coherent states~\eqref{eq:coh}, the probability of getting a ``click'' in the bin $n_z$ is  given by the sum of encompassing probabilities:
\beq
\pr(n_z|\Psi_N)= \sum_{\frac{j_z}{N}\in [ -\frac{1}{2} + \frac{n_z-1}{N_{b}}, -\frac{1}{2} + \frac{n_z}{N_{b}} [} \pr(j_z|\Psi_N).
\eeq
In the limit of large $N$, the continuous approximation of this probability reads:
\beq
\pr(n_z|\Psi_N) \approxeq \smashoperator{\int_{-\frac{1}{2} + \frac{n_z-1}{N_{b}}}^{-\frac{1}{2} + \frac{n_z}{N_{b}}}} d \left(\frac{j_z}{N}\right) \frac{1}{\sqrt{2\pi \Delta^{2}(\tilde{Z}_N|\Psi_N)}} e^{-\frac{\left(\frac{j_{z}}{N}-\<\tilde{Z}_N\>\right)^{2}}{2 \Delta^{2}(\tilde{Z}_N|\Psi_N)}}.     
\eeq
Remembering that $\Delta^{2}(\tilde{Z}_N|\Psi_N) = p(1-p)/N$, it is clear that the distribution $\pr(n_z|\Psi_N)$ will also concentrate around the value $\<\tilde{Z}_N\>=1/2-p$. Differently from before, however, the number of outcomes is fixed (expressed by integration limits independent of $N$). This means $\pr(n_z|\Psi_N) $ will concentrate around the bin that contains $j_z/N=p-1/2$. Such a bin is  $\floor{N_b p}+1$ for $0\le p <1$, and $N_b$ for $p=1$. All the other bins will have probabilities decreasing exponentially with $N$.

\begin{figure}
	\centering
	\includegraphics[width=\linewidth]{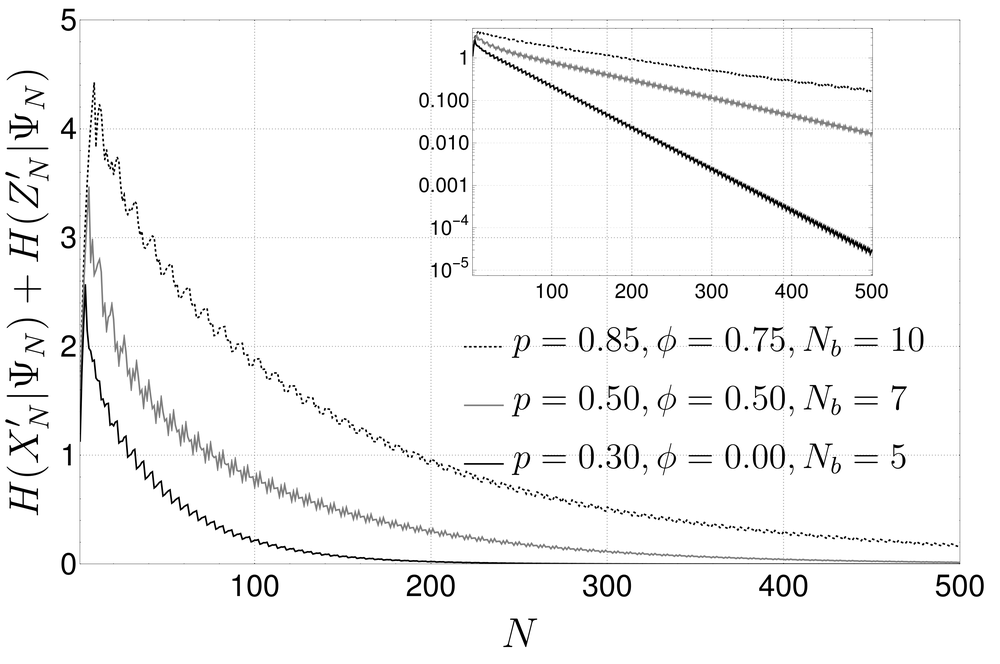}  
	\caption{\textbf{Sum of entropies for increasing system size.} As the number of constituents increases, it becomes possible for coherent-spin states to have well-defined magnetization simultaneously in $x$ and $z$ directions. An increase in the sum of entropies is observed while the number of spins is smaller than the number of bins. After this initial period, an exponential-like decay is established (see inset). The small oscillation present in the curves is due to the ratio between $N_b$ and $N$. \label{fig:systemsize}}	
\end{figure}

\begin{figure}
	\centering
	\includegraphics[width=\linewidth]{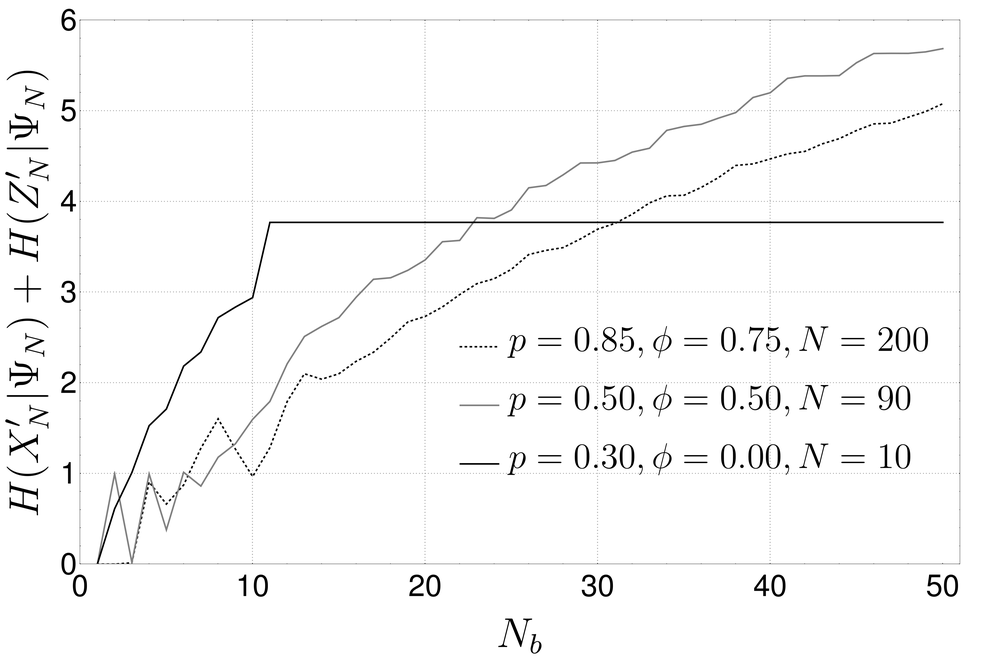}  
	\caption{\textbf{Sum of entropies for increasing number of bins.} As the number of bins increases their width shrinks and we can access finer details of the system. For $N_b \gtrsim N$ the preparation uncertainty is not negligible, and quantum features are sizeable. Notice that for $N_b> N$ each possible value of magnetization will be in a different bin, and there will bins that do not contain any possible value. From that point on the sum of entropies is constant, as seen for the black solid case.
		\label{fig:binsize}}
	
\end{figure}

In the limit of  $N\rightarrow \infty$ the distribution  $\pr(n_z|\Psi_N)$ will thus tend to a delta function fully contained in a single bin. In this way the entropy $H(Z^\prime_N|\Psi_{N})=- \sum_{n_z=1}^{N_b} \pr(n_z|\Psi_N) \log \pr(n_z|\Psi_N)$ will vanish. A completely analogous argument shows that $H(X^\prime_N|\Psi_{N})$ will also vanish in the macroscopic limit. We then recover the classically expected behavior:
\beq
\lim_{N\rightarrow \infty} H(X^\prime_N|\Psi_{N})+H(Z^\prime_N|\Psi_{N})=0.
\eeq
The recovery of this  classical signature is numerically observed in Figs.\ref{fig:systemsize} and \ref{fig:binsize}. As in Ref.~\cite{toscano}, for finite $N$ the sum of entropies will be always greater than zero. Nevertheless this deviation won't be visible for macroscopic systems. Pathological cases, where the sum of entropies won't vanish, are when either $\<\tilde{X}_N\>$ or $\<\tilde{Z}_N\>$ are exactly equal to values separating two contiguous bins. These cases, however, are of zero volume and will never occur in real experiments.

Lastly, note that a similar classical behaviour would be obtained even if we increased the number of bins with the system size, but not faster than $\sqrt{N}$. That is because the variance of the probability in \eqref{eq:totalspin} concentrates around the mean as $1/\sqrt{N}$. Thus, if the number of outcomes grows slower than the distribution concentrates, the distribution will eventually  be contained within a single bin and the entropy will vanish.


\section{Conclusion} Uncertainty relations are one of the cornerstones of quantum mechanics. Since its introduction by Heisenberg, the possibility of preparing a system with well-defined properties was linked to the commutation relation between the associated observables. It is only with the advent of quantum information techniques that a more clear cut understanding of the classical limit of these relations is now possible. 

Differently from what was described by von Neumann~\cite{neumann29}, an effective commutation is not necessary to recover a classical behaviour. Notice that $X^\prime_N$ and $Z^\prime_N$ do not commute for any system size.  Notably, we find that  it is only by including imprecision in the macroscopic observables that a classical character is recovered. Similar conclusions were achieved in Ref.~\cite{oleg2020} for the scenario of consecutive coarse-grained measurements.

Moreover, from the above results it is also clear that if quantum properties are desirable even in large systems, like in large quantum computers, the number of outcomes in preparation measurements has to grow faster than  $\sqrt{N}$.


\begin{acknowledgements}
 We would like to thank Daniel Schneider for the question that lead to these results,  Yelena Guryanova for comments on an early draft, and Roberto Sarthour for discussions on NMR measurements.
This work is supported by the Brazilian funding agencies
CNPq and CAPES, and it is part of the Brazilian National
Institute for Quantum Information.
\end{acknowledgements}
%
%

\bibliographystyle{apsrev4-1}	

\begin{thebibliography}{50}%
	\makeatletter
	\providecommand \@ifxundefined [1]{%
		\@ifx{#1\undefined}
	}%
	\providecommand \@ifnum [1]{%
		\ifnum #1\expandafter \@firstoftwo
		\else \expandafter \@secondoftwo
		\fi
	}%
	\providecommand \@ifx [1]{%
		\ifx #1\expandafter \@firstoftwo
		\else \expandafter \@secondoftwo
		\fi
	}%
	\providecommand \natexlab [1]{#1}%
	\providecommand \enquote  [1]{``#1''}%
	\providecommand \bibnamefont  [1]{#1}%
	\providecommand \bibfnamefont [1]{#1}%
	\providecommand \citenamefont [1]{#1}%
	\providecommand \href@noop [0]{\@secondoftwo}%
	\providecommand \href [0]{\begingroup \@sanitize@url \@href}%
	\providecommand \@href[1]{\@@startlink{#1}\@@href}%
	\providecommand \@@href[1]{\endgroup#1\@@endlink}%
	\providecommand \@sanitize@url [0]{\catcode `\\12\catcode `\$12\catcode
		`\&12\catcode `\#12\catcode `\^12\catcode `\_12\catcode `\%12\relax}%
	\providecommand \@@startlink[1]{}%
	\providecommand \@@endlink[0]{}%
	\providecommand \url  [0]{\begingroup\@sanitize@url \@url }%
	\providecommand \@url [1]{\endgroup\@href {#1}{\urlprefix }}%
	\providecommand \urlprefix  [0]{URL }%
	\providecommand \Eprint [0]{\href }%
	\providecommand \doibase [0]{http://dx.doi.org/}%
	\providecommand \selectlanguage [0]{\@gobble}%
	\providecommand \bibinfo  [0]{\@secondoftwo}%
	\providecommand \bibfield  [0]{\@secondoftwo}%
	\providecommand \translation [1]{[#1]}%
	\providecommand \BibitemOpen [0]{}%
	\providecommand \bibitemStop [0]{}%
	\providecommand \bibitemNoStop [0]{.\EOS\space}%
	\providecommand \EOS [0]{\spacefactor3000\relax}%
	\providecommand \BibitemShut  [1]{\csname bibitem#1\endcsname}%
	\let\auto@bib@innerbib\@empty
	\bibitem [{\citenamefont {Mandelstam}\ and\ \citenamefont
		{Leontowitsch}(1928)}]{MandelstamTunneling}%
	\BibitemOpen
	\bibfield  {author} {\bibinfo {author} {\bibfnamefont {L.}~\bibnamefont
			{Mandelstam}}\ and\ \bibinfo {author} {\bibfnamefont {M.}~\bibnamefont
			{Leontowitsch}},\ }\href@noop {} {\bibfield  {journal} {\bibinfo  {journal}
			{Zeitschrift f{\"u}r Physik}\ }\textbf {\bibinfo {volume} {47}},\ \bibinfo
		{pages} {131} (\bibinfo {year} {1928})}\BibitemShut {NoStop}%
	\bibitem [{\citenamefont {Einstein}\ \emph {et~al.}(1935)\citenamefont
		{Einstein}, \citenamefont {Podolsky},\ and\ \citenamefont {Rosen}}]{EPR}%
	\BibitemOpen
	\bibfield  {author} {\bibinfo {author} {\bibfnamefont {A.}~\bibnamefont
			{Einstein}}, \bibinfo {author} {\bibfnamefont {B.}~\bibnamefont {Podolsky}},
		\ and\ \bibinfo {author} {\bibfnamefont {N.}~\bibnamefont {Rosen}},\
	}\href@noop {} {\bibfield  {journal} {\bibinfo  {journal} {Physical review}\
		}\textbf {\bibinfo {volume} {47}},\ \bibinfo {pages} {777} (\bibinfo {year}
		{1935})}\BibitemShut {NoStop}%
	\bibitem [{\citenamefont
		{Schr{\"o}dinger}(1935{\natexlab{a}})}]{schroedinger1935}%
	\BibitemOpen
	\bibfield  {author} {\bibinfo {author} {\bibfnamefont {E.}~\bibnamefont
			{Schr{\"o}dinger}},\ }in\ \href@noop {} {\emph {\bibinfo {booktitle}
			{Mathematical Proceedings of the Cambridge Philosophical Society}}},\
	Vol.~\bibinfo {volume} {31}\ (\bibinfo {organization} {Cambridge University
		Press},\ \bibinfo {year} {1935})\ pp.\ \bibinfo {pages}
	{555--563}\BibitemShut {NoStop}%
	\bibitem [{\citenamefont {Horodecki}\ \emph {et~al.}(2009)\citenamefont
		{Horodecki}, \citenamefont {Horodecki}, \citenamefont {Horodecki},\ and\
		\citenamefont {Horodecki}}]{HorodeckiRev}%
	\BibitemOpen
	\bibfield  {author} {\bibinfo {author} {\bibfnamefont {R.}~\bibnamefont
			{Horodecki}}, \bibinfo {author} {\bibfnamefont {P.}~\bibnamefont
			{Horodecki}}, \bibinfo {author} {\bibfnamefont {M.}~\bibnamefont
			{Horodecki}}, \ and\ \bibinfo {author} {\bibfnamefont {K.}~\bibnamefont
			{Horodecki}},\ }\href {\doibase 10.1103/RevModPhys.81.865} {\bibfield
		{journal} {\bibinfo  {journal} {Rev. Mod. Phys.}\ }\textbf {\bibinfo {volume}
			{81}},\ \bibinfo {pages} {865} (\bibinfo {year} {2009})}\BibitemShut
	{NoStop}%
	\bibitem [{\citenamefont
		{Schr{\"o}dinger}(1935{\natexlab{b}})}]{SchrodingerCat}%
	\BibitemOpen
	\bibfield  {author} {\bibinfo {author} {\bibfnamefont {E.}~\bibnamefont
			{Schr{\"o}dinger}},\ }\href@noop {} {\bibfield  {journal} {\bibinfo
			{journal} {Naturwissenschaften}\ }\textbf {\bibinfo {volume} {23}},\ \bibinfo
		{pages} {823} (\bibinfo {year} {1935}{\natexlab{b}})}\BibitemShut {NoStop}%
	\bibitem [{\citenamefont {Caldeira}\ and\ \citenamefont
		{Leggett}(1981)}]{caldeira}%
	\BibitemOpen
	\bibfield  {author} {\bibinfo {author} {\bibfnamefont {A.~O.}\ \bibnamefont
			{Caldeira}}\ and\ \bibinfo {author} {\bibfnamefont {A.~J.}\ \bibnamefont
			{Leggett}},\ }\href {\doibase 10.1103/PhysRevLett.46.211} {\bibfield
		{journal} {\bibinfo  {journal} {Phys. Rev. Lett.}\ }\textbf {\bibinfo
			{volume} {46}},\ \bibinfo {pages} {211} (\bibinfo {year} {1981})}\BibitemShut
	{NoStop}%
	\bibitem [{\citenamefont {Zurek}(2003{\natexlab{a}})}]{zurek2003}%
	\BibitemOpen
	\bibfield  {author} {\bibinfo {author} {\bibfnamefont {W.~H.}\ \bibnamefont
			{Zurek}},\ }\href@noop {} {\bibfield  {journal} {\bibinfo  {journal} {arXiv
				preprint quant-ph/0306072}\ } (\bibinfo {year}
		{2003}{\natexlab{a}})}\BibitemShut {NoStop}%
	\bibitem [{\citenamefont {Schlosshauer}(2007)}]{schlosshauer}%
	\BibitemOpen
	\bibfield  {author} {\bibinfo {author} {\bibfnamefont {M.~A.}\ \bibnamefont
			{Schlosshauer}},\ }\href@noop {} {\emph {\bibinfo {title} {Decoherence: and
				the quantum-to-classical transition}}}\ (\bibinfo  {publisher} {Springer
		Science \& Business Media},\ \bibinfo {year} {2007})\BibitemShut {NoStop}%
	\bibitem [{\citenamefont {Eibenberger}\ \emph {et~al.}(2013)\citenamefont
		{Eibenberger}, \citenamefont {Gerlich}, \citenamefont {Arndt}, \citenamefont
		{Mayor},\ and\ \citenamefont {T{\"u}xen}}]{Eibenberger}%
	\BibitemOpen
	\bibfield  {author} {\bibinfo {author} {\bibfnamefont {S.}~\bibnamefont
			{Eibenberger}}, \bibinfo {author} {\bibfnamefont {S.}~\bibnamefont
			{Gerlich}}, \bibinfo {author} {\bibfnamefont {M.}~\bibnamefont {Arndt}},
		\bibinfo {author} {\bibfnamefont {M.}~\bibnamefont {Mayor}}, \ and\ \bibinfo
		{author} {\bibfnamefont {J.}~\bibnamefont {T{\"u}xen}},\ }\href@noop {}
	{\bibfield  {journal} {\bibinfo  {journal} {Physical Chemistry Chemical
				Physics}\ }\textbf {\bibinfo {volume} {15}},\ \bibinfo {pages} {14696}
		(\bibinfo {year} {2013})}\BibitemShut {NoStop}%
	\bibitem [{\citenamefont {O’Connell}\ \emph {et~al.}(2010)\citenamefont
		{O’Connell}, \citenamefont {Hofheinz}, \citenamefont {Ansmann},
		\citenamefont {Bialczak}, \citenamefont {Lenander}, \citenamefont {Lucero},
		\citenamefont {Neeley}, \citenamefont {Sank}, \citenamefont {Wang},
		\citenamefont {Weides} \emph {et~al.}}]{connell}%
	\BibitemOpen
	\bibfield  {author} {\bibinfo {author} {\bibfnamefont {A.~D.}\ \bibnamefont
			{O’Connell}}, \bibinfo {author} {\bibfnamefont {M.}~\bibnamefont
			{Hofheinz}}, \bibinfo {author} {\bibfnamefont {M.}~\bibnamefont {Ansmann}},
		\bibinfo {author} {\bibfnamefont {R.~C.}\ \bibnamefont {Bialczak}}, \bibinfo
		{author} {\bibfnamefont {M.}~\bibnamefont {Lenander}}, \bibinfo {author}
		{\bibfnamefont {E.}~\bibnamefont {Lucero}}, \bibinfo {author} {\bibfnamefont
			{M.}~\bibnamefont {Neeley}}, \bibinfo {author} {\bibfnamefont
			{D.}~\bibnamefont {Sank}}, \bibinfo {author} {\bibfnamefont {H.}~\bibnamefont
			{Wang}}, \bibinfo {author} {\bibfnamefont {M.}~\bibnamefont {Weides}},  \emph
		{et~al.},\ }\href@noop {} {\bibfield  {journal} {\bibinfo  {journal}
			{Nature}\ }\textbf {\bibinfo {volume} {464}},\ \bibinfo {pages} {697}
		(\bibinfo {year} {2010})}\BibitemShut {NoStop}%
	\bibitem [{\citenamefont {Riedinger}\ \emph {et~al.}(2018)\citenamefont
		{Riedinger}, \citenamefont {Wallucks}, \citenamefont {Marinkovi{\'c}},
		\citenamefont {L{\"o}schnauer}, \citenamefont {Aspelmeyer}, \citenamefont
		{Hong},\ and\ \citenamefont {Gr{\"o}blacher}}]{riedinger}%
	\BibitemOpen
	\bibfield  {author} {\bibinfo {author} {\bibfnamefont {R.}~\bibnamefont
			{Riedinger}}, \bibinfo {author} {\bibfnamefont {A.}~\bibnamefont {Wallucks}},
		\bibinfo {author} {\bibfnamefont {I.}~\bibnamefont {Marinkovi{\'c}}},
		\bibinfo {author} {\bibfnamefont {C.}~\bibnamefont {L{\"o}schnauer}},
		\bibinfo {author} {\bibfnamefont {M.}~\bibnamefont {Aspelmeyer}}, \bibinfo
		{author} {\bibfnamefont {S.}~\bibnamefont {Hong}}, \ and\ \bibinfo {author}
		{\bibfnamefont {S.}~\bibnamefont {Gr{\"o}blacher}},\ }\href@noop {}
	{\bibfield  {journal} {\bibinfo  {journal} {Nature}\ }\textbf {\bibinfo
			{volume} {556}},\ \bibinfo {pages} {473} (\bibinfo {year}
		{2018})}\BibitemShut {NoStop}%
	\bibitem [{\citenamefont {Havl{\'\i}{\v{c}}ek}\ \emph
		{et~al.}(2019)\citenamefont {Havl{\'\i}{\v{c}}ek}, \citenamefont
		{C{\'o}rcoles}, \citenamefont {Temme}, \citenamefont {Harrow}, \citenamefont
		{Kandala}, \citenamefont {Chow},\ and\ \citenamefont {Gambetta}}]{ibm}%
	\BibitemOpen
	\bibfield  {author} {\bibinfo {author} {\bibfnamefont {V.}~\bibnamefont
			{Havl{\'\i}{\v{c}}ek}}, \bibinfo {author} {\bibfnamefont {A.~D.}\
			\bibnamefont {C{\'o}rcoles}}, \bibinfo {author} {\bibfnamefont
			{K.}~\bibnamefont {Temme}}, \bibinfo {author} {\bibfnamefont {A.~W.}\
			\bibnamefont {Harrow}}, \bibinfo {author} {\bibfnamefont {A.}~\bibnamefont
			{Kandala}}, \bibinfo {author} {\bibfnamefont {J.~M.}\ \bibnamefont {Chow}}, \
		and\ \bibinfo {author} {\bibfnamefont {J.~M.}\ \bibnamefont {Gambetta}},\
	}\href@noop {} {\bibfield  {journal} {\bibinfo  {journal} {Nature}\ }\textbf
		{\bibinfo {volume} {567}},\ \bibinfo {pages} {209} (\bibinfo {year}
		{2019})}\BibitemShut {NoStop}%
	\bibitem [{\citenamefont {Wright}\ \emph {et~al.}(2019)\citenamefont {Wright},
		\citenamefont {Beck}, \citenamefont {Debnath}, \citenamefont {Amini},
		\citenamefont {Nam}, \citenamefont {Grzesiak}, \citenamefont {Chen},
		\citenamefont {Pisenti}, \citenamefont {Chmielewski}, \citenamefont {Collins}
		\emph {et~al.}}]{ionq}%
	\BibitemOpen
	\bibfield  {author} {\bibinfo {author} {\bibfnamefont {K.}~\bibnamefont
			{Wright}}, \bibinfo {author} {\bibfnamefont {K.}~\bibnamefont {Beck}},
		\bibinfo {author} {\bibfnamefont {S.}~\bibnamefont {Debnath}}, \bibinfo
		{author} {\bibfnamefont {J.}~\bibnamefont {Amini}}, \bibinfo {author}
		{\bibfnamefont {Y.}~\bibnamefont {Nam}}, \bibinfo {author} {\bibfnamefont
			{N.}~\bibnamefont {Grzesiak}}, \bibinfo {author} {\bibfnamefont {J.-S.}\
			\bibnamefont {Chen}}, \bibinfo {author} {\bibfnamefont {N.}~\bibnamefont
			{Pisenti}}, \bibinfo {author} {\bibfnamefont {M.}~\bibnamefont
			{Chmielewski}}, \bibinfo {author} {\bibfnamefont {C.}~\bibnamefont
			{Collins}},  \emph {et~al.},\ }\href {\doibase 10.1038/s41467-019-13534-2}
	{\bibfield  {journal} {\bibinfo  {journal} {Nat. Commun.}\ }\textbf {\bibinfo
			{volume} {10}},\ \bibinfo {pages} {1} (\bibinfo {year} {2019})}\BibitemShut
	{NoStop}%
	\bibitem [{\citenamefont {Arute}\ \emph {et~al.}(2019)\citenamefont {Arute},
		\citenamefont {Arya}, \citenamefont {Babbush}, \citenamefont {Bacon},
		\citenamefont {Bardin}, \citenamefont {Barends}, \citenamefont {Biswas},
		\citenamefont {Boixo}, \citenamefont {Brandao}, \citenamefont {Buell} \emph
		{et~al.}}]{google}%
	\BibitemOpen
	\bibfield  {author} {\bibinfo {author} {\bibfnamefont {F.}~\bibnamefont
			{Arute}}, \bibinfo {author} {\bibfnamefont {K.}~\bibnamefont {Arya}},
		\bibinfo {author} {\bibfnamefont {R.}~\bibnamefont {Babbush}}, \bibinfo
		{author} {\bibfnamefont {D.}~\bibnamefont {Bacon}}, \bibinfo {author}
		{\bibfnamefont {J.~C.}\ \bibnamefont {Bardin}}, \bibinfo {author}
		{\bibfnamefont {R.}~\bibnamefont {Barends}}, \bibinfo {author} {\bibfnamefont
			{R.}~\bibnamefont {Biswas}}, \bibinfo {author} {\bibfnamefont
			{S.}~\bibnamefont {Boixo}}, \bibinfo {author} {\bibfnamefont {F.~G.}\
			\bibnamefont {Brandao}}, \bibinfo {author} {\bibfnamefont {D.~A.}\
			\bibnamefont {Buell}},  \emph {et~al.},\ }\href@noop {} {\bibfield  {journal}
		{\bibinfo  {journal} {Nature}\ }\textbf {\bibinfo {volume} {574}},\ \bibinfo
		{pages} {505} (\bibinfo {year} {2019})}\BibitemShut {NoStop}%
	\bibitem [{\citenamefont {Mermin}(1980)}]{mermin1980}%
	\BibitemOpen
	\bibfield  {author} {\bibinfo {author} {\bibfnamefont {N.~D.}\ \bibnamefont
			{Mermin}},\ }\href {\doibase 10.1103/PhysRevD.22.356} {\bibfield  {journal}
		{\bibinfo  {journal} {Phys. Rev. D}\ }\textbf {\bibinfo {volume} {22}},\
		\bibinfo {pages} {356} (\bibinfo {year} {1980})}\BibitemShut {NoStop}%
	\bibitem [{\citenamefont {Poulin}(2005)}]{poulin2005}%
	\BibitemOpen
	\bibfield  {author} {\bibinfo {author} {\bibfnamefont {D.}~\bibnamefont
			{Poulin}},\ }\href@noop {} {\bibfield  {journal} {\bibinfo  {journal}
			{Physical Review A}\ }\textbf {\bibinfo {volume} {71}},\ \bibinfo {pages}
		{022102} (\bibinfo {year} {2005})}\BibitemShut {NoStop}%
	\bibitem [{\citenamefont {Kofler}\ and\ \citenamefont
		{Brukner}(2008)}]{caslavLG}%
	\BibitemOpen
	\bibfield  {author} {\bibinfo {author} {\bibfnamefont {J.}~\bibnamefont
			{Kofler}}\ and\ \bibinfo {author} {\bibfnamefont {{\v{C}}.}~\bibnamefont
			{Brukner}},\ }\href@noop {} {\bibfield  {journal} {\bibinfo  {journal}
			{Physical review letters}\ }\textbf {\bibinfo {volume} {101}},\ \bibinfo
		{pages} {090403} (\bibinfo {year} {2008})}\BibitemShut {NoStop}%
	\bibitem [{\citenamefont {Raeisi}\ \emph {et~al.}(2011)\citenamefont {Raeisi},
		\citenamefont {Sekatski},\ and\ \citenamefont {Simon}}]{Raeisi2011}%
	\BibitemOpen
	\bibfield  {author} {\bibinfo {author} {\bibfnamefont {S.}~\bibnamefont
			{Raeisi}}, \bibinfo {author} {\bibfnamefont {P.}~\bibnamefont {Sekatski}}, \
		and\ \bibinfo {author} {\bibfnamefont {C.}~\bibnamefont {Simon}},\ }\href
	{\doibase 10.1103/PhysRevLett.107.250401} {\bibfield  {journal} {\bibinfo
			{journal} {Phys. Rev. Lett.}\ }\textbf {\bibinfo {volume} {107}},\ \bibinfo
		{pages} {250401} (\bibinfo {year} {2011})}\BibitemShut {NoStop}%
	\bibitem [{\citenamefont {Wang}\ \emph {et~al.}(2013)\citenamefont {Wang},
		\citenamefont {Ghobadi}, \citenamefont {Raeisi},\ and\ \citenamefont
		{Simon}}]{Wang2013}%
	\BibitemOpen
	\bibfield  {author} {\bibinfo {author} {\bibfnamefont {T.}~\bibnamefont
			{Wang}}, \bibinfo {author} {\bibfnamefont {R.}~\bibnamefont {Ghobadi}},
		\bibinfo {author} {\bibfnamefont {S.}~\bibnamefont {Raeisi}}, \ and\ \bibinfo
		{author} {\bibfnamefont {C.}~\bibnamefont {Simon}},\ }\href {\doibase
		10.1103/PhysRevA.88.062114} {\bibfield  {journal} {\bibinfo  {journal} {Phys.
				Rev. A}\ }\textbf {\bibinfo {volume} {88}},\ \bibinfo {pages} {062114}
		(\bibinfo {year} {2013})}\BibitemShut {NoStop}%
	\bibitem [{\citenamefont {Jeong}\ \emph {et~al.}(2014)\citenamefont {Jeong},
		\citenamefont {Lim},\ and\ \citenamefont {Kim}}]{Jeong2014}%
	\BibitemOpen
	\bibfield  {author} {\bibinfo {author} {\bibfnamefont {H.}~\bibnamefont
			{Jeong}}, \bibinfo {author} {\bibfnamefont {Y.}~\bibnamefont {Lim}}, \ and\
		\bibinfo {author} {\bibfnamefont {M.~S.}\ \bibnamefont {Kim}},\ }\href
	{\doibase 10.1103/PhysRevLett.112.010402} {\bibfield  {journal} {\bibinfo
			{journal} {Phys. Rev. Lett.}\ }\textbf {\bibinfo {volume} {112}},\ \bibinfo
		{pages} {010402} (\bibinfo {year} {2014})}\BibitemShut {NoStop}%
	\bibitem [{\citenamefont {Park}\ \emph {et~al.}(2014)\citenamefont {Park},
		\citenamefont {Ji}, \citenamefont {Lee},\ and\ \citenamefont
		{Nha}}]{Park2014}%
	\BibitemOpen
	\bibfield  {author} {\bibinfo {author} {\bibfnamefont {J.}~\bibnamefont
			{Park}}, \bibinfo {author} {\bibfnamefont {S.-W.}\ \bibnamefont {Ji}},
		\bibinfo {author} {\bibfnamefont {J.}~\bibnamefont {Lee}}, \ and\ \bibinfo
		{author} {\bibfnamefont {H.}~\bibnamefont {Nha}},\ }\href {\doibase
		10.1103/PhysRevA.89.042102} {\bibfield  {journal} {\bibinfo  {journal} {Phys.
				Rev. A}\ }\textbf {\bibinfo {volume} {89}},\ \bibinfo {pages} {042102}
		(\bibinfo {year} {2014})}\BibitemShut {NoStop}%
	\bibitem [{\citenamefont {Duarte}\ \emph {et~al.}(2017)\citenamefont {Duarte},
		\citenamefont {Carvalho}, \citenamefont {Bernardes},\ and\ \citenamefont
		{de~Melo}}]{cris2017}%
	\BibitemOpen
	\bibfield  {author} {\bibinfo {author} {\bibfnamefont {C.}~\bibnamefont
			{Duarte}}, \bibinfo {author} {\bibfnamefont {G.~D.}\ \bibnamefont
			{Carvalho}}, \bibinfo {author} {\bibfnamefont {N.~K.}\ \bibnamefont
			{Bernardes}}, \ and\ \bibinfo {author} {\bibfnamefont {F.}~\bibnamefont
			{de~Melo}},\ }\href {\doibase 10.1103/PhysRevA.96.032113} {\bibfield
		{journal} {\bibinfo  {journal} {Phys. Rev. A}\ }\textbf {\bibinfo {volume}
			{96}},\ \bibinfo {pages} {032113} (\bibinfo {year} {2017})}\BibitemShut
	{NoStop}%
	\bibitem [{\citenamefont {Silva~Correia}\ and\ \citenamefont
		{de~Melo}(2019)}]{pedrinho}%
	\BibitemOpen
	\bibfield  {author} {\bibinfo {author} {\bibfnamefont {P.}~\bibnamefont
			{Silva~Correia}}\ and\ \bibinfo {author} {\bibfnamefont {F.}~\bibnamefont
			{de~Melo}},\ }\href {\doibase 10.1103/PhysRevA.100.022334} {\bibfield
		{journal} {\bibinfo  {journal} {Phys. Rev. A}\ }\textbf {\bibinfo {volume}
			{100}},\ \bibinfo {pages} {022334} (\bibinfo {year} {2019})}\BibitemShut
	{NoStop}%
	\bibitem [{\citenamefont {Kabernik}(2018)}]{oleg}%
	\BibitemOpen
	\bibfield  {author} {\bibinfo {author} {\bibfnamefont {O.}~\bibnamefont
			{Kabernik}},\ }\href {\doibase 10.1103/PhysRevA.97.052130} {\bibfield
		{journal} {\bibinfo  {journal} {Phys. Rev. A}\ }\textbf {\bibinfo {volume}
			{97}},\ \bibinfo {pages} {052130} (\bibinfo {year} {2018})}\BibitemShut
	{NoStop}%
	\bibitem [{\citenamefont {Duarte}(2019)}]{cris2019}%
	\BibitemOpen
	\bibfield  {author} {\bibinfo {author} {\bibfnamefont {C.}~\bibnamefont
			{Duarte}},\ }\href@noop {} {\bibfield  {journal} {\bibinfo  {journal} {arXiv
				preprint arXiv:1908.04432}\ } (\bibinfo {year} {2019})}\BibitemShut {NoStop}%
	\bibitem [{\citenamefont {Zurek}(2003{\natexlab{b}})}]{ZurekRMP}%
	\BibitemOpen
	\bibfield  {author} {\bibinfo {author} {\bibfnamefont {W.~H.}\ \bibnamefont
			{Zurek}},\ }\href {\doibase 10.1103/RevModPhys.75.715} {\bibfield  {journal}
		{\bibinfo  {journal} {Rev. Mod. Phys.}\ }\textbf {\bibinfo {volume} {75}},\
		\bibinfo {pages} {715} (\bibinfo {year} {2003}{\natexlab{b}})}\BibitemShut
	{NoStop}%
	\bibitem [{\citenamefont {Alicki}\ \emph {et~al.}(2009)\citenamefont {Alicki},
		\citenamefont {Fannes},\ and\ \citenamefont {Pogorzelska}}]{alicki2009}%
	\BibitemOpen
	\bibfield  {author} {\bibinfo {author} {\bibfnamefont {R.}~\bibnamefont
			{Alicki}}, \bibinfo {author} {\bibfnamefont {M.}~\bibnamefont {Fannes}}, \
		and\ \bibinfo {author} {\bibfnamefont {M.}~\bibnamefont {Pogorzelska}},\
	}\href {\doibase 10.1103/PhysRevA.79.052111} {\bibfield  {journal} {\bibinfo
			{journal} {Phys. Rev. A}\ }\textbf {\bibinfo {volume} {79}},\ \bibinfo
		{pages} {052111} (\bibinfo {year} {2009})}\BibitemShut {NoStop}%
	\bibitem [{\citenamefont {Kabernik}\ \emph {et~al.}(2020)\citenamefont
		{Kabernik}, \citenamefont {Pollack},\ and\ \citenamefont
		{Singh}}]{oleg2020a}%
	\BibitemOpen
	\bibfield  {author} {\bibinfo {author} {\bibfnamefont {O.}~\bibnamefont
			{Kabernik}}, \bibinfo {author} {\bibfnamefont {J.}~\bibnamefont {Pollack}}, \
		and\ \bibinfo {author} {\bibfnamefont {A.}~\bibnamefont {Singh}},\ }\href
	{\doibase 10.1103/PhysRevA.101.032303} {\bibfield  {journal} {\bibinfo
			{journal} {Phys. Rev. A}\ }\textbf {\bibinfo {volume} {101}},\ \bibinfo
		{pages} {032303} (\bibinfo {year} {2020})}\BibitemShut {NoStop}%
	\bibitem [{\citenamefont {Ozawa}(2003)}]{Ozawa03}%
	\BibitemOpen
	\bibfield  {author} {\bibinfo {author} {\bibfnamefont {M.}~\bibnamefont
			{Ozawa}},\ }\href {\doibase 10.1103/PhysRevA.67.042105} {\bibfield  {journal}
		{\bibinfo  {journal} {Phys. Rev. A}\ }\textbf {\bibinfo {volume} {67}},\
		\bibinfo {pages} {042105} (\bibinfo {year} {2003})}\BibitemShut {NoStop}%
	\bibitem [{\citenamefont {Busch}\ \emph {et~al.}(2013)\citenamefont {Busch},
		\citenamefont {Lahti},\ and\ \citenamefont {Werner}}]{Busch13}%
	\BibitemOpen
	\bibfield  {author} {\bibinfo {author} {\bibfnamefont {P.}~\bibnamefont
			{Busch}}, \bibinfo {author} {\bibfnamefont {P.}~\bibnamefont {Lahti}}, \ and\
		\bibinfo {author} {\bibfnamefont {R.~F.}\ \bibnamefont {Werner}},\ }\href
	{\doibase 10.1103/PhysRevLett.111.160405} {\bibfield  {journal} {\bibinfo
			{journal} {Phys. Rev. Lett.}\ }\textbf {\bibinfo {volume} {111}},\ \bibinfo
		{pages} {160405} (\bibinfo {year} {2013})}\BibitemShut {NoStop}%
	\bibitem [{\citenamefont {Heisenberg}(1985)}]{HeisenbergUncertainty}%
	\BibitemOpen
	\bibfield  {author} {\bibinfo {author} {\bibfnamefont {W.}~\bibnamefont
			{Heisenberg}},\ }in\ \href@noop {} {\emph {\bibinfo {booktitle} {Original
				Scientific Papers Wissenschaftliche Originalarbeiten}}}\ (\bibinfo
	{publisher} {Springer},\ \bibinfo {year} {1985})\ pp.\ \bibinfo {pages}
	{478--504}\BibitemShut {NoStop}%
	\bibitem [{\citenamefont {Robertson}(1929)}]{Robertson}%
	\BibitemOpen
	\bibfield  {author} {\bibinfo {author} {\bibfnamefont {H.~P.}\ \bibnamefont
			{Robertson}},\ }\href@noop {} {\bibfield  {journal} {\bibinfo  {journal}
			{Physical Review}\ }\textbf {\bibinfo {volume} {34}},\ \bibinfo {pages} {163}
		(\bibinfo {year} {1929})}\BibitemShut {NoStop}%
	\bibitem [{\citenamefont {Neumann}(1929)}]{neumann29}%
	\BibitemOpen
	\bibfield  {author} {\bibinfo {author} {\bibfnamefont {J.~v.}\ \bibnamefont
			{Neumann}},\ }\href@noop {} {\bibfield  {journal} {\bibinfo  {journal}
			{Zeitschrift f{\"u}r Physik}\ }\textbf {\bibinfo {volume} {57}},\ \bibinfo
		{pages} {30} (\bibinfo {year} {1929})}\BibitemShut {NoStop}%
	\bibitem [{\citenamefont {Deutsch}(1983)}]{Deutsch1983}%
	\BibitemOpen
	\bibfield  {author} {\bibinfo {author} {\bibfnamefont {D.}~\bibnamefont
			{Deutsch}},\ }\href@noop {} {\bibfield  {journal} {\bibinfo  {journal}
			{Physical Review Letters}\ }\textbf {\bibinfo {volume} {50}},\ \bibinfo
		{pages} {631} (\bibinfo {year} {1983})}\BibitemShut {NoStop}%
	\bibitem [{\citenamefont {Kraus}(1987)}]{Kraus}%
	\BibitemOpen
	\bibfield  {author} {\bibinfo {author} {\bibfnamefont {K.}~\bibnamefont
			{Kraus}},\ }\href@noop {} {\bibfield  {journal} {\bibinfo  {journal}
			{Physical Review D}\ }\textbf {\bibinfo {volume} {35}},\ \bibinfo {pages}
		{3070} (\bibinfo {year} {1987})}\BibitemShut {NoStop}%
	\bibitem [{\citenamefont {Maassen}\ and\ \citenamefont
		{Uffink}(1988)}]{Maassen}%
	\BibitemOpen
	\bibfield  {author} {\bibinfo {author} {\bibfnamefont {H.}~\bibnamefont
			{Maassen}}\ and\ \bibinfo {author} {\bibfnamefont {J.~B.}\ \bibnamefont
			{Uffink}},\ }\href@noop {} {\bibfield  {journal} {\bibinfo  {journal}
			{Physical Review Letters}\ }\textbf {\bibinfo {volume} {60}},\ \bibinfo
		{pages} {1103} (\bibinfo {year} {1988})}\BibitemShut {NoStop}%
	\bibitem [{\citenamefont {Wehner}\ and\ \citenamefont
		{Winter}(2010)}]{review1}%
	\BibitemOpen
	\bibfield  {author} {\bibinfo {author} {\bibfnamefont {S.}~\bibnamefont
			{Wehner}}\ and\ \bibinfo {author} {\bibfnamefont {A.}~\bibnamefont
			{Winter}},\ }\href@noop {} {\bibfield  {journal} {\bibinfo  {journal} {New
				Journal of Physics}\ }\textbf {\bibinfo {volume} {12}},\ \bibinfo {pages}
		{025009} (\bibinfo {year} {2010})}\BibitemShut {NoStop}%
	\bibitem [{\citenamefont {Toscano}\ \emph {et~al.}(2018)\citenamefont
		{Toscano}, \citenamefont {Tasca}, \citenamefont {Rudnicki},\ and\
		\citenamefont {Walborn}}]{review2}%
	\BibitemOpen
	\bibfield  {author} {\bibinfo {author} {\bibfnamefont {F.}~\bibnamefont
			{Toscano}}, \bibinfo {author} {\bibfnamefont {D.~S.}\ \bibnamefont {Tasca}},
		\bibinfo {author} {\bibfnamefont {{\L}.}~\bibnamefont {Rudnicki}}, \ and\
		\bibinfo {author} {\bibfnamefont {S.~P.}\ \bibnamefont {Walborn}},\
	}\href@noop {} {\bibfield  {journal} {\bibinfo  {journal} {Entropy}\ }\textbf
		{\bibinfo {volume} {20}},\ \bibinfo {pages} {454} (\bibinfo {year}
		{2018})}\BibitemShut {NoStop}%
	\bibitem [{\citenamefont {Coles}\ \emph {et~al.}(2017)\citenamefont {Coles},
		\citenamefont {Berta}, \citenamefont {Tomamichel},\ and\ \citenamefont
		{Wehner}}]{review3}%
	\BibitemOpen
	\bibfield  {author} {\bibinfo {author} {\bibfnamefont {P.~J.}\ \bibnamefont
			{Coles}}, \bibinfo {author} {\bibfnamefont {M.}~\bibnamefont {Berta}},
		\bibinfo {author} {\bibfnamefont {M.}~\bibnamefont {Tomamichel}}, \ and\
		\bibinfo {author} {\bibfnamefont {S.}~\bibnamefont {Wehner}},\ }\href
	{\doibase 10.1103/RevModPhys.89.015002} {\bibfield  {journal} {\bibinfo
			{journal} {Rev. Mod. Phys.}\ }\textbf {\bibinfo {volume} {89}},\ \bibinfo
		{pages} {015002} (\bibinfo {year} {2017})}\BibitemShut {NoStop}%
	\bibitem [{\citenamefont {Bia{\l}ynicki-Birula}\ and\ \citenamefont
		{Mycielski}(1975)}]{Birula1975}%
	\BibitemOpen
	\bibfield  {author} {\bibinfo {author} {\bibfnamefont {I.}~\bibnamefont
			{Bia{\l}ynicki-Birula}}\ and\ \bibinfo {author} {\bibfnamefont
			{J.}~\bibnamefont {Mycielski}},\ }\href@noop {} {\bibfield  {journal}
		{\bibinfo  {journal} {Communications in Mathematical Physics}\ }\textbf
		{\bibinfo {volume} {44}},\ \bibinfo {pages} {129} (\bibinfo {year}
		{1975})}\BibitemShut {NoStop}%
	\bibitem [{\citenamefont {Glauber}(1963)}]{glauber}%
	\BibitemOpen
	\bibfield  {author} {\bibinfo {author} {\bibfnamefont {R.~J.}\ \bibnamefont
			{Glauber}},\ }\href {\doibase 10.1103/PhysRev.131.2766} {\bibfield  {journal}
		{\bibinfo  {journal} {Phys. Rev.}\ }\textbf {\bibinfo {volume} {131}},\
		\bibinfo {pages} {2766} (\bibinfo {year} {1963})}\BibitemShut {NoStop}%
	\bibitem [{\citenamefont {Arecchi}\ \emph {et~al.}(1972)\citenamefont
		{Arecchi}, \citenamefont {Courtens}, \citenamefont {Gilmore},\ and\
		\citenamefont {Thomas}}]{arecchi}%
	\BibitemOpen
	\bibfield  {author} {\bibinfo {author} {\bibfnamefont {F.}~\bibnamefont
			{Arecchi}}, \bibinfo {author} {\bibfnamefont {E.}~\bibnamefont {Courtens}},
		\bibinfo {author} {\bibfnamefont {R.}~\bibnamefont {Gilmore}}, \ and\
		\bibinfo {author} {\bibfnamefont {H.}~\bibnamefont {Thomas}},\ }\href@noop {}
	{\bibfield  {journal} {\bibinfo  {journal} {Physical Review A}\ }\textbf
		{\bibinfo {volume} {6}},\ \bibinfo {pages} {2211} (\bibinfo {year}
		{1972})}\BibitemShut {NoStop}%
	\bibitem [{\citenamefont {Zurek}(2009)}]{zurek}%
	\BibitemOpen
	\bibfield  {author} {\bibinfo {author} {\bibfnamefont {W.~H.}\ \bibnamefont
			{Zurek}},\ }\href@noop {} {\bibfield  {journal} {\bibinfo  {journal} {Nature
				Physics}\ }\textbf {\bibinfo {volume} {5}},\ \bibinfo {pages} {181} (\bibinfo
		{year} {2009})}\BibitemShut {NoStop}%
	\bibitem [{\citenamefont {Brandao}\ \emph {et~al.}(2015)\citenamefont
		{Brandao}, \citenamefont {Piani},\ and\ \citenamefont {Horodecki}}]{brandao}%
	\BibitemOpen
	\bibfield  {author} {\bibinfo {author} {\bibfnamefont {F.~G.}\ \bibnamefont
			{Brandao}}, \bibinfo {author} {\bibfnamefont {M.}~\bibnamefont {Piani}}, \
		and\ \bibinfo {author} {\bibfnamefont {P.}~\bibnamefont {Horodecki}},\
	}\href@noop {} {\bibfield  {journal} {\bibinfo  {journal} {Nature
				communications}\ }\textbf {\bibinfo {volume} {6}},\ \bibinfo {pages} {7908}
		(\bibinfo {year} {2015})}\BibitemShut {NoStop}%
	\bibitem [{\citenamefont {Oliveira}\ \emph {et~al.}(2019)\citenamefont
		{Oliveira}, \citenamefont {de~Paula},\ and\ \citenamefont
		{Drumond}}]{sheilla}%
	\BibitemOpen
	\bibfield  {author} {\bibinfo {author} {\bibfnamefont {S.~M.}\ \bibnamefont
			{Oliveira}}, \bibinfo {author} {\bibfnamefont {A.~L.}\ \bibnamefont
			{de~Paula}}, \ and\ \bibinfo {author} {\bibfnamefont {R.~C.}\ \bibnamefont
			{Drumond}},\ }\href {\doibase 10.1103/PhysRevA.100.052110} {\bibfield
		{journal} {\bibinfo  {journal} {Phys. Rev. A}\ }\textbf {\bibinfo {volume}
			{100}},\ \bibinfo {pages} {052110} (\bibinfo {year} {2019})}\BibitemShut
	{NoStop}%
	\bibitem [{\citenamefont {Leggett}\ and\ \citenamefont {Garg}(1985)}]{LG}%
	\BibitemOpen
	\bibfield  {author} {\bibinfo {author} {\bibfnamefont {A.~J.}\ \bibnamefont
			{Leggett}}\ and\ \bibinfo {author} {\bibfnamefont {A.}~\bibnamefont {Garg}},\
	}\href {\doibase 10.1103/PhysRevLett.54.857} {\bibfield  {journal} {\bibinfo
			{journal} {Phys. Rev. Lett.}\ }\textbf {\bibinfo {volume} {54}},\ \bibinfo
		{pages} {857} (\bibinfo {year} {1985})}\BibitemShut {NoStop}%
	\bibitem [{\citenamefont {Feller}(2008)}]{feller}%
	\BibitemOpen
	\bibfield  {author} {\bibinfo {author} {\bibfnamefont {W.}~\bibnamefont
			{Feller}},\ }\href@noop {} {\emph {\bibinfo {title} {An introduction to
				probability theory and its applications}}},\ Vol.~\bibinfo {volume} {2}\
	(\bibinfo  {publisher} {John Wiley \& Sons},\ \bibinfo {year}
	{2008})\BibitemShut {NoStop}%
	\bibitem [{\citenamefont {Oliveira}\ \emph {et~al.}(2011)\citenamefont
		{Oliveira}, \citenamefont {Sarthour~Jr}, \citenamefont {Bonagamba},
		\citenamefont {Azevedo},\ and\ \citenamefont {Freitas}}]{NMRbook}%
	\BibitemOpen
	\bibfield  {author} {\bibinfo {author} {\bibfnamefont {I.}~\bibnamefont
			{Oliveira}}, \bibinfo {author} {\bibfnamefont {R.}~\bibnamefont
			{Sarthour~Jr}}, \bibinfo {author} {\bibfnamefont {T.}~\bibnamefont
			{Bonagamba}}, \bibinfo {author} {\bibfnamefont {E.}~\bibnamefont {Azevedo}},
		\ and\ \bibinfo {author} {\bibfnamefont {J.~C.}\ \bibnamefont {Freitas}},\
	}\href@noop {} {\emph {\bibinfo {title} {NMR quantum information
				processing}}}\ (\bibinfo  {publisher} {Elsevier},\ \bibinfo {year}
	{2011})\BibitemShut {NoStop}%
	\bibitem [{\citenamefont {Rudnicki}\ \emph {et~al.}(2012)\citenamefont
		{Rudnicki}, \citenamefont {Walborn},\ and\ \citenamefont
		{Toscano}}]{toscano}%
	\BibitemOpen
	\bibfield  {author} {\bibinfo {author} {\bibfnamefont {{\L}.}~\bibnamefont
			{Rudnicki}}, \bibinfo {author} {\bibfnamefont {S.~P.}\ \bibnamefont
			{Walborn}}, \ and\ \bibinfo {author} {\bibfnamefont {F.}~\bibnamefont
			{Toscano}},\ }\href@noop {} {\bibfield  {journal} {\bibinfo  {journal}
			{Physical Review A}\ }\textbf {\bibinfo {volume} {85}},\ \bibinfo {pages}
		{042115} (\bibinfo {year} {2012})}\BibitemShut {NoStop}%
	\bibitem [{\citenamefont {Kabernik}(2020)}]{oleg2020}%
	\BibitemOpen
	\bibfield  {author} {\bibinfo {author} {\bibfnamefont {O.}~\bibnamefont
			{Kabernik}},\ }\href@noop {} {\bibfield  {journal} {\bibinfo  {journal}
			{arXiv preprint arXiv:2002.01564}\ } (\bibinfo {year} {2020})}\BibitemShut
	{NoStop}%
\end{thebibliography}
%
\end{document}